\documentclass[12pt]{article}
\usepackage{graphicx}
\usepackage{amssymb}
\newlength{\extraspace}
\newlength{\extraspaces}
\newcommand{\be}{\begin{equation}
\addtolength{\abovedisplayskip}{\extraspaces}
\addtolength{\belowdisplayskip}{\extraspaces}
\addtolength{\abovedisplayshortskip}{\extraspace}
\addtolength{\belowdisplayshortskip}{\extraspace}}
\newcommand{\ee}{\end{equation}}

\newcommand{\ba}{\begin{eqnarray}
\addtolength{\abovedisplayskip}{\extraspaces}
\addtolength{\belowdisplayskip}{\extraspaces}
\addtolength{\abovedisplayshortskip}{\extraspace}
\addtolength{\belowdisplayshortskip}{\extraspace}}
\newcommand{\ea}{\end{eqnarray}}

\newcommand{\newsection}[1]{
\vspace{12mm}
\pagebreak[3]
\addtocounter{section}{1}
\setcounter{equation}{0}
\setcounter{subsection}{0}
\noindent{\bf \thesection. #1}
\nopagebreak
\medskip
\nopagebreak}

\newcommand{\newsubsection}[1]{
\vspace{0.8cm}
\pagebreak[3]
\addtocounter{subsection}{1}
\noindent{\it \thesubsection. #1}
\nopagebreak
\vspace{2mm}
\nopagebreak}

\newcounter{saveeqn}

\newcommand{\dif}{\mathrm{d}}

\begin{document}
\addtolength{\baselineskip}{1.5mm}

\thispagestyle{empty}
\begin{flushright}

\end{flushright}
\vbox{}
\vspace{2cm}

\begin{center}
{\LARGE{A five-parameter class of solutions to the vacuum \\[2mm]
Einstein equations 
        }}\\[16mm]
{{Yu Chen~~and~~Edward Teo}}
\\[6mm]
{\it Department of Physics,
National University of Singapore, 
Singapore 119260}\\[15mm]

\end{center}
\vspace{2cm}

\centerline{\bf Abstract}
\bigskip
\noindent
We present a new five-parameter class of Ricci-flat solutions in four dimensions with Euclidean signature. The solution is asymptotically locally flat (ALF), and contains a finite asymptotic NUT charge. When this charge is sent to infinity, the solution becomes asymptotically locally Euclidean (ALE), and one in fact obtains the Ricci-flat Pleba\'nski--Demia\'nski solution. The solution we have found can thus be regarded as an ALF generalisation of the latter solution. We also show that it can be interpreted as a system consisting of two touching Kerr-NUTs: the south pole of one Kerr-NUT touches the north pole of the other. The total NUT charge of such a system is then identified with the asymptotic NUT charge. Setting the asymptotic NUT charge to zero gives a four-parameter asymptotically flat (AF) solution, and contained within this subclass is the completely regular two-parameter AF instanton previously discovered by the present authors. Various other limits are also discussed, including that of the triple-collinearly-centered Gibbons--Hawking solution, and an ALF generalisation of the C-metric.


\newpage

\newsection{Introduction}

Exact solutions have played an important role in the development of Einstein's general theory of relativity, describing important predictions such as black holes. The first known exact solution was discovered by Schwarzschild in 1918, although it was only realised much later that it describes a static black hole. The rotating generalisation of the Schwarzschild solution was discovered by Kerr in 1963. It is a two-parameter class of solutions, describing a rotating mass in general relativity.

The Schwarzschild solution has another well-known generalisation, the so-called Taub-NUT solution \cite{Newman:1963}. It possesses some peculiar properties, such as the presence of a so-called Misner string. Indeed, the Taub-NUT solution was used by Misner as ``a counterexample to almost anything'' \cite{Misner}. When a rotational parameter is added, it generalises to the Kerr-NUT solution, which is contained within the so-called Carter--Pleba\'nski solution \cite{Carter,Plebanski}. The latter solution is more general, containing electric and magnetic charges, as well as a cosmological constant. But in the present paper, we will only be interested in its Ricci-flat limit, i.e., the Kerr-NUT solution.

A black hole can also accelerate in general relativity. The solution that describes such a black hole is known as the C-metric. The C-metric was found quite early on, but its interpretation as an accelerating black hole was only made clear after the work of Kinnersley and Walker \cite{Kinnersley:1970zw} in 1970. Kinnersley also found its generalisation with rotation and NUT charge \cite{Kinnersley:1969zza}. In 1976, Pleba\'nski and Demia\'nski obtained the most general black hole solution, and presented it in a remarkably compact form \cite{Plebanski:1976gy}. It contains seven parameters: mass, rotation, NUT charge, acceleration, cosmological constant, electric and magnetic charges. Restricting to the Ricci-flat class, the cosmological constant and electric and magnetic charges are removed. We shall call such a class of solutions the Ricci-flat Pleba\'nski--Demia\'nski solution, although it is perhaps more appropriate to call it the Kinnersley solution. It is a four-parameter class of solutions.

In this paper, we focus on solutions with Euclidean signature, i.e., those with all-plus signature. Interestingly, all the above-mentioned black-hole solutions have Euclidean sections. Many calculations in the black-hole space-times, in particular those involving quantum fields, are first done in their Euclidean sections and then analytically continued back to Lorentzian signature. Euclidean solutions are of interest in their own right as well. In the Euclidean path integral approach to quantum gravity \cite{Gibbons:1994}, all the Euclidean metrics on a given manifold with fixed boundary conditions are integrated over. Those metrics satisfying the Einstein equations are thus the stationary phase points of the path integral.

We should, however, point out that it is not guaranteed that a Euclidean solution can be obtained from a Lorentzian one by Wick rotation, or {\it vice versa\/}. For example, the self-dual class of Euclidean solutions known as the multi-Taub-NUT solution in the Gibbons--Hawking ansatz \cite{Hawking:1976jb,Gibbons:1979zt} has no Lorentzian section. Nevertheless, as mentioned, large classes of solutions do have Euclidean sections, and these will be our main focus from now on. From this point, we will implicitly refer to its Euclidean section when we talk about a solution. For example, when we refer to the Kerr-NUT solution, we really mean the Euclidean Kerr-NUT solution.

In the Euclidean section, the time-translational symmetry becomes either a translational or a rotational symmetry. The Euclidean spaces now possess two commuting Killing vectors, one corresponding to the Euclidean time flow, and the other to the usual rotation. The black-hole horizons, as well as the so-called acceleration horizons when the acceleration parameter is present, become axes of the space. This means that some linear combination of the two Killing vector fields vanishes at each of these horizons or axes. The classification of Euclidean solutions based on the fixed-point sets of the Killing vector fields was carried out by Gibbons and Hawking \cite{Gibbons:1979c}, and more recently in terms of the so-called rod structure by the present authors \cite{Chen:2010zu}.

In the rod-structure formalism, the various axes of the space are known as rods. Each rod has a direction, which is defined to be the normalised Killing vector field which vanishes along that rod. The points at which adjacent rods meet are called turning points. The reader is referred to \cite{Chen:2010zu}, and references therein, for more details of the rod-structure formalism, and for explicit examples of rod structures of a number of Euclidean solutions.

Euclidean solutions can also be classified by their behaviour at asymptotic infinity as being asymptotically flat (AF), asymptotically locally flat (ALF), asymptotically Euclidean (AE) or asymptotically locally Euclidean (ALE) \cite{Gibbons:1979gd}. A solution is said to be AF, if at infinity the metric approaches the form:
\be
\label{definition_AF}
\dif s^2_{\rm AF}\equiv \dif \tau^2+\dif r^2+r^2(\dif \theta^2+\sin^2\theta\,\dif {\phi}^2)\,,
\ee
sufficiently fast. Solutions in this class include Schwarzschild and Kerr. On the other hand, if the metric approaches
\be
\label{definition_ALF}
\dif s^2_{\rm ALF}\equiv \left(\dif {\tau}+2n\cos\theta\,\dif{\phi}\right)^2+\dif r^2+r^2(\dif \theta^2+\sin^2\theta\,\dif {\phi}^2)\,,
\ee
sufficiently fast at infinity, it is said to be ALF. Here the so-called NUT charge $n$ manifests itself in the asymptotic behaviour of this class of metrics. Solutions of this type include Taub-NUT and Kerr-NUT. If the metric at infinity instead approaches
\be
\label{definition_ALE}
\dif s^2_{\rm ALE}\equiv r^2\cos^2\theta\,\dif\tau^2+\dif r^2+r^2(\dif \theta^2+\sin^2\theta\,\dif {\phi}^2)\,,
\ee
sufficiently fast, it is called AE or ALE, depending on whether the asymptotic constant $r$-surface is identified. We refer to these two classes as ALE collectively when global geometry is of no concern; in fact, AE is just a special case of ALE with a trivial identification group. Flat space is of course AE. A well-known example of an ALE space is the double-centered Gibbons--Hawking space, with the Eguchi--Hanson space as a special case.

If one is only interested in local geometry, AF metrics are a special limit of ALF metrics with \emph{vanishing} NUT charge. This is clear from the asymptotics (\ref{definition_AF}) and (\ref{definition_ALF}). On the other hand, AE and ALE metrics are a special limit of ALF metrics with \emph{infinite} NUT charge. The latter correspondence is not manifest in the asymptotics (\ref{definition_ALF}) and (\ref{definition_ALE}) alone, but it is true for all explicitly known solutions (c.f.~Table 1). For example, the infinite NUT charge limit of Taub-NUT gives flat space. In this sense, ALF metrics are the most general class of solutions.

\begin{table}[!t]
\begin{center}
\begin{tabular}{|c|c|c|}
  \hline
{\bf ALF metric} & {\bf ALE limit} & {\bf Number of} \\
 &  & {\bf turning points} \\
\hline
Taub-NUT &  flat space & 1\\
\hline
&  double-centered & \\
Kerr-NUT &  Gibbons--Hawking & 2\\
\hline
$n$-centered & $n$-centered & \\
Taub-NUT & Gibbons--Hawking & $n$ \\
\hline
\end{tabular}
\caption{Known examples of ALE metrics as special limits of ALF metrics with infinite NUT charge. The last column refers to the number of turning points in the rod structures of the corresponding ALF and ALE metrics.}
\end{center}
\end{table}

It turns out that the C-metric is ALE: the acceleration horizon now becomes one of the asymptotic axes in the Euclidean section. The more general Ricci-flat Pleba\'nski--Demia\'nski solution is also ALE, as can be checked explicitly. One is then naturally led to the question of whether the latter metric admits an ALF generalisation. Such a generalisation, if it exists, would contain an asymptotic NUT charge $n$, such that when $n$ is taken to infinity, the Ricci-flat Pleba\'nski--Demia\'nski solution is recovered.

The aim of this paper is to explicitly present this ALF generalisation of the Ricci-flat Pleba\'nski--Demia\'nski solution, and to analyse its properties and various limits. Like the Ricci-flat Pleba\'nski--Demia\'nski solution, this new solution has three turning points in its rod structure. It can be regarded as the non-self-dual counterpart of the triple-collinearly-centered Taub-NUT solution, just as the Kerr-NUT solution is the non-self-dual counterpart of the double-centered Taub-NUT solution in the two-turning-point case. In analogy with the Kerr-NUT solution, we will see that this new solution in fact contains the self-dual triple-collinearly-centered Gibbons--Hawking solution as another ALE limit.

It should be emphasised that the asymptotic NUT charge $n$ present in the new ALF solution that we have found, is different from the NUT-charge parameter that the Pleba\'nski--Demia\'nski solution is known to possess. In particular, in the limit when $n$ is taken to infinity, the latter parameter is still present in the solution and remains finite. The difference between these two types of NUT charges can be understood in terms of the rod structure of the solution. It turns out that the asymptotic NUT charge is a property relating the two asymptotic rods (the first and fourth rods of the rod structure), while the traditional NUT charge is an analogous property relating the first and third rods.\footnote{The NUT-charge parameter in the Pleba\'nski--Demia\'nski solution is so called because it reduces to the NUT-charge parameter in the Kerr-NUT solution when the zero-acceleration limit is taken. In this limit, the fourth rod disappears from the rod structure, and the first and third rods become the asymptotic rods. Thus, it is only in this limit that the NUT-charge parameter in the Pleba\'nski--Demia\'nski solution has its usual interpretation as in (\ref{definition_ALF}).}

The paper is organised as follows. We begin in Sec.~2 by presenting the general metric together with an analysis of its symmetries. We also briefly discuss the construction of the solution. The main geometrical properties of the solution are analysed in Sec.~3. This includes the conditions necessary to ensure the correct metric signature and absence of curvature singularities, as well as its asymptotic structure and rod structure. In Sec.~4, we show how four important limits of the solution can be obtained. They all share the property that the number of turning points in the rod structure remains fixed at three. In Sec.~5, we provide an alternative interpretation of the solution as a system consisting of two touching Kerr-NUTs, while in Sec.~6, we discuss the interpretation of this solution in Kaluza--Klein theory as a system of three collinearly centered monopoles. The paper ends with a discussion of possible generalisations and future problems. There is also an appendix, which contains details of how the two- and one-turning-point limits of the solution are taken.

\newsection{The metric}

The solution we have found can be compactly written in the following form:
\ba
\label{metric_general}
\dif s^2&=&\frac{\left(F\dif \tau+G\dif \phi\right)^2}{(x-y)HF}+\frac{kH}{(x-y)^3}\left(\frac{\dif x^2}{X}-\frac{\dif y^2}{Y}-\frac{XY}{kF}\,\dif \phi^2\right),\cr
H&=&(\nu x+ y)[(\nu x-y)(a_1-a_3xy)-2(1-\nu)(a_0-a_4x^2y^2)]\,,\quad F=y^2X-x^2Y\,,\cr
G&=&(\nu^2a_0+2\nu a_3y^3+2\nu a_4y^4-a_4y^4)X+(a_0-2\nu a_0-2\nu a_1x-\nu^2 a_4x^4)Y\,,\cr
X&=&a_0+a_1x+a_2x^2+a_3x^3+a_4x^4,\quad Y=a_0+a_1y+a_2y^2+a_3y^3+a_4y^4,
\ea
in the coordinate system ($\tau,\phi,x,y$). There are seven apparent parameters in this solution: an overall scale factor $k$, a parameter $\nu$, and five arbitrary coefficients $a_{0,\dots,4}$ for the quartic polynomial $X$ (or $Y$).

\newsubsection{Symmetries}

It can be checked that the solution (\ref{metric_general}) has the following two continuous symmetries:

\begin{itemize}
\item Scaling symmetry:
\be
\label{symmetry_scaling}
a_i\rightarrow a_i/c^i,\quad x\rightarrow cx\,,\quad y\rightarrow cy\,,\quad \phi\rightarrow c^2\phi\,;
\ee
\item Parameter symmetry:
\be
a_i\rightarrow ca_i\,,\quad \phi\rightarrow \phi/c\,,
\ee
for an arbitrary non-zero constant $c$, and $i=0,\dots,4$.
\end{itemize}
This implies that two out of the five coefficients $a_{0,\dots,4}$ are actually redundant. So (\ref{metric_general}) is in fact a five-parameter solution. There are also two discrete symmetries present:

\begin{itemize}
\item Swapping symmetry:
\be
\nu\rightarrow1/\nu\,,\quad x\leftrightarrow y\,,\quad (\tau,\phi)\rightarrow i(\tau/\nu-2a_2\phi,-\nu\phi)\,,\quad k\rightarrow -k\nu^2,
\ee
where $i$ is the imaginary unit;
\item Inversion-swapping symmetry:
\be
x\rightarrow\frac{1}{y}\,,\quad y\rightarrow\frac{1}{x}\,, \quad a_0\leftrightarrow a_4\,,\quad a_1\leftrightarrow a_3\,.
\ee
\end{itemize}

One can use these symmetries to narrow the ranges of the coordinates and parameters of the solution. For example, the swapping symmetry can be used to restrict the range of $\nu$ to $-1\leq \nu\leq 1$ without any loss of generality. This will be used in our discussion below.

\newsubsection{Alternative parameterisation}

It is often convenient to reparameterise the solution in terms of the roots of the polynomial $X$ as
\be
\label{alt_param}
X=a_4(x-x_1)(x-x_2)(x-x_3)(x-x_4)\,,
\ee
by defining
\ba
a_0&=&a_4x_1x_2x_3x_4\,,\quad a_2=a_4(x_1x_2+x_1x_3+x_1x_4+x_2x_3+x_2x_4+x_3x_4)\,,\cr
a_1&=&-a_4(x_1x_2x_3+x_1x_2x_4+x_1x_3x_4+x_2x_3x_4)\,,\quad a_3=-a_4(x_1+x_2+x_3+x_4)\,.
\ea
Then the scaling symmetry implies that the following substitutions:
\be
a_4\rightarrow a_4/c^4\,,\quad x_i\rightarrow cx_i\,,\quad x\rightarrow cx\,,\quad y\rightarrow cy\,,\quad \phi\rightarrow c^2\phi,
\ee
is a symmetry, so the relevant parameters are the relative ratios, rather than the definite values, of the roots of $X$. The parameter symmetry implies that the quartic coefficient $a_4$ of $X$ can be set to an arbitrary non-zero value.

\newsubsection{ISM construction}

The solution was constructed using the inverse-scattering method (ISM) \cite{Belinski:2001,Pomeransky:2005}. It was originally obtained by applying a three-soliton transformation on the Euclidean triple-collinearly-centered Schwarzschild solution. The three so-called Belinski--Zakharov (BZ) parameters were then fixed to eliminate the three corresponding turning points where the soliton-transformation was performed. One possible choice of these BZ parameters then led to the above solution (the other choice leads to the triple-collinearly-centered Taub-NUT solution). It was subsequently realised that this solution can also be generated by applying a three-soliton transformation on the Euclidean double-Schwarzschild solution, and then eliminating one turning point by fixing the corresponding BZ parameter. The latter construction was in fact previously carried out in \cite{Chen:2011tc}, from which the new AF gravitational instanton was extracted as a special case.

The remaining major challenge was to then cast the solution in a compact form. We eventually succeeded by using the C-metric-like coordinates $(x,y)$. The basic idea was to absorb as many parameters as possible into the structure function $X$. A M\"obius transformation can be performed on $(x,y)$, and using this freedom we fixed one factor of $H$ as $\nu x+y$.

The details of the above ISM constructions and simplifications will be presented elsewhere.

\newsection{Analysis of the geometry}

\newsubsection{Ranges of coordinates and parameters}

To simplify our analysis, we need to narrow the ranges and remove certain redundancies of the parameters. We use the swapping symmetry to restrict the range of $\nu$, and choose a negative $x_2$ using the scaling symmetry:
\be
\label{parameter_range1}
-1\leq \nu\leq 1\,,\quad x_2<0\,.
\ee
These ranges shall be assumed throughout the paper. $x_2$ can be further gauge-fixed to be $-1$ due to the scaling symmetry, and $a_4$ can be fixed to be $1$ due to the parameter symmetry:
\be
\label{parameter_range2}
x_2=-1\,,\quad a_4=1\,.
\ee
We shall use these gauge-fixed values whenever it is convenient.

We now deduce the appropriate ranges of the coordinates $(x,y)$. To ensure that the signature of the space does not change in the region of interest, we note that $x$ must lie between a pair of adjacent roots of $X$, and similarly for $y$. Furthermore, these two pairs of adjacent roots must be such that $X$ has opposite sign to $Y$, to ensure a positive- or negative-definite signature. In particular, this implies that $x$ and $y$ lie in different ranges.

We also require that the region of interest is non-compact with an asymptotic infinity. As can be seen from the metric (\ref{metric_general}), asymptotic infinity is reached when $x=y$. Given that $x$ and $y$ necessarily lie in different ranges, there must exist a common boundary at which they coincide. Without loss of generality, we take this common boundary to be at $x=y=x_2$. We further assume that $x$ ranges between the two roots $x_2$ and $x_3$, and that $y$ ranges between the two roots $x_1$ and $x_2$.

At this stage, we note that the region of interest can be visualised as a rectangle or ``box'' in a two-dimensional plot with $x$ and $y$ as the axes. This box will have four sides, corresponding to the boundaries of the ranges of $x$ and $y$. Asymptotic infinity itself is represented by the diagonal line $x=y$ in this plot. The box will then have to touch the point $x=y=x_2$ on this line at either its lower-right or upper-left corner.

Depending on the values of $x_1$ and $x_3$ relative to $x_2$, the box may either be finite in extent, or up to two of its sides may extend to infinity. For example, consider a box touching the point $x=y=x_2$ at its lower-right corner. If we have the ordering $x_3<x_2<x_1$, the box will be finite in extent. On the other hand, if $x_2<x_3$, the left side of the box will extend to infinity and ``wrap around'' to the right side of the plot, ending at $x=x_3$. Similarly, if $x_1<x_2$, the top side of the box will extend to infinity and wrap around to the bottom side of the plot, ending at $y=x_1$. We thus have four qualitatively different cases: (i) $x_3<x_2<x_1$; (ii) $x_2<x_3$ and $x_2<x_1$; (iii) $x_3<x_2$ and $x_1<x_2$; and (iv) $x_1<x_2<x_3$.

We next turn to the constraints set by requiring that the region of interest is free of curvature singularities. We note that there are possible curvature singularities whenever $H$ vanishes. As can be seen from the expression of $H$ in (\ref{metric_general}), this occurs along the straight line $y=-\nu x$, as well as on non-trivial curves where the second factor of $H$ vanishes. We also note that there are curvature singularities at the four points $(x=0,y=\pm \infty)$ and $(x=\pm\infty,y=0)$.

We now show that a box touching the point $x=y=x_2$ at its lower-right corner will always contain a curvature singularity. As explained above, there are four different cases to consider, labelled by (i)--(iv). It turns out that cases (iii) and (iv) can immediately be eliminated, since we must have $x_2<x_1<1$ in order for the box not to touch the line singularity $y=-\nu x$ for $-1\leq \nu\leq 1$. Turning to case (ii), note that we require $x_1<0$ in order for the box not to contain the singularity at $(x=-\infty,y=0)$. If this holds, then case (ii) is in fact equivalent to case (i) under the inversion-swapping symmetry. Thus we are left with case (i) to consider. It can be checked that in this case, $H$ will necessarily change sign somewhere in the box. Indeed, if one demands that the box avoids the line singularity $y=-\nu x$, it will inevitably contain part of the non-trivial singularity curves.

\begin{figure}[!t]
\begin{center}
\includegraphics[height=3.1in,width=3.1in]{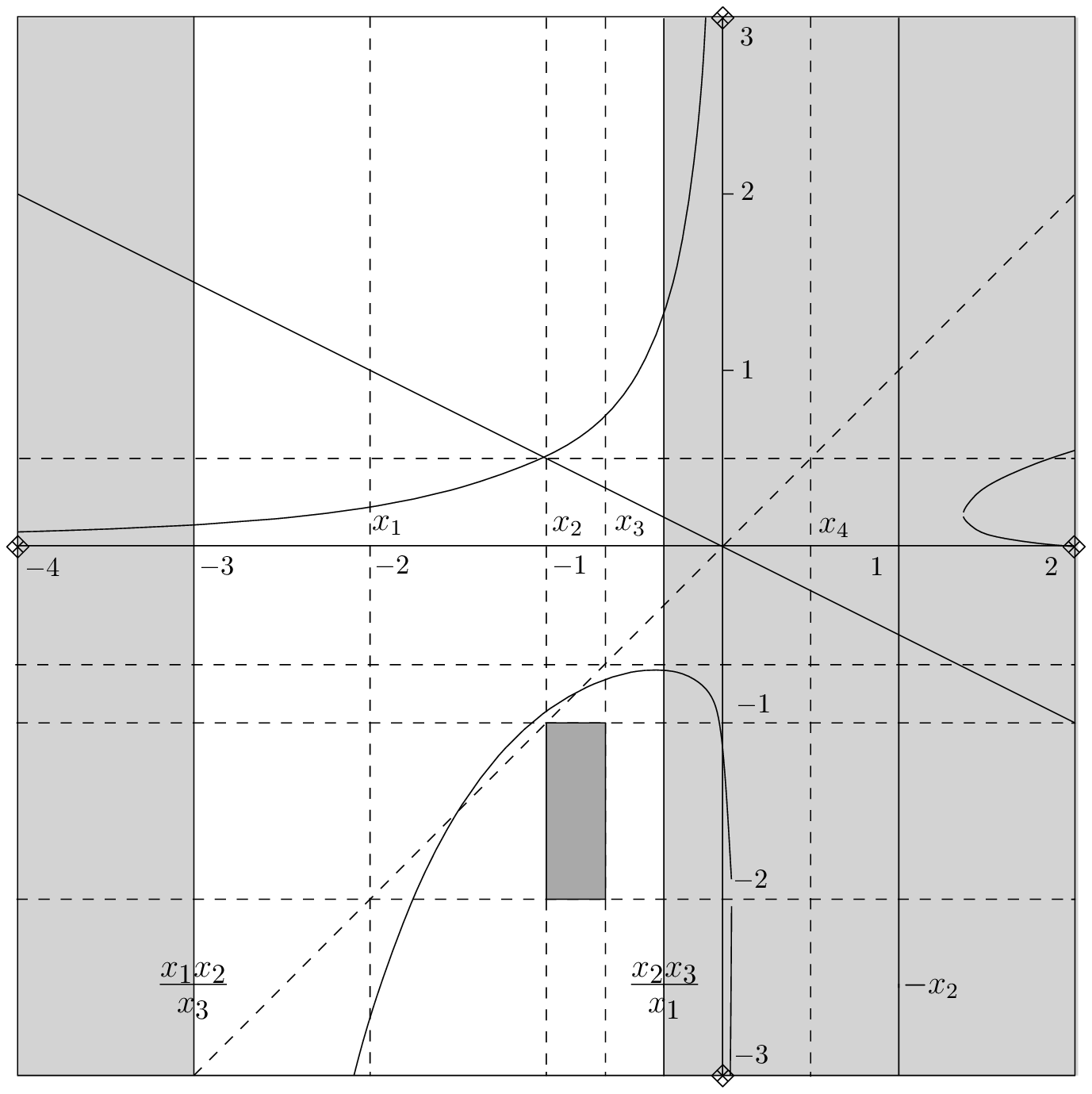}
~~~\includegraphics[height=3.1in,width=3.1in]{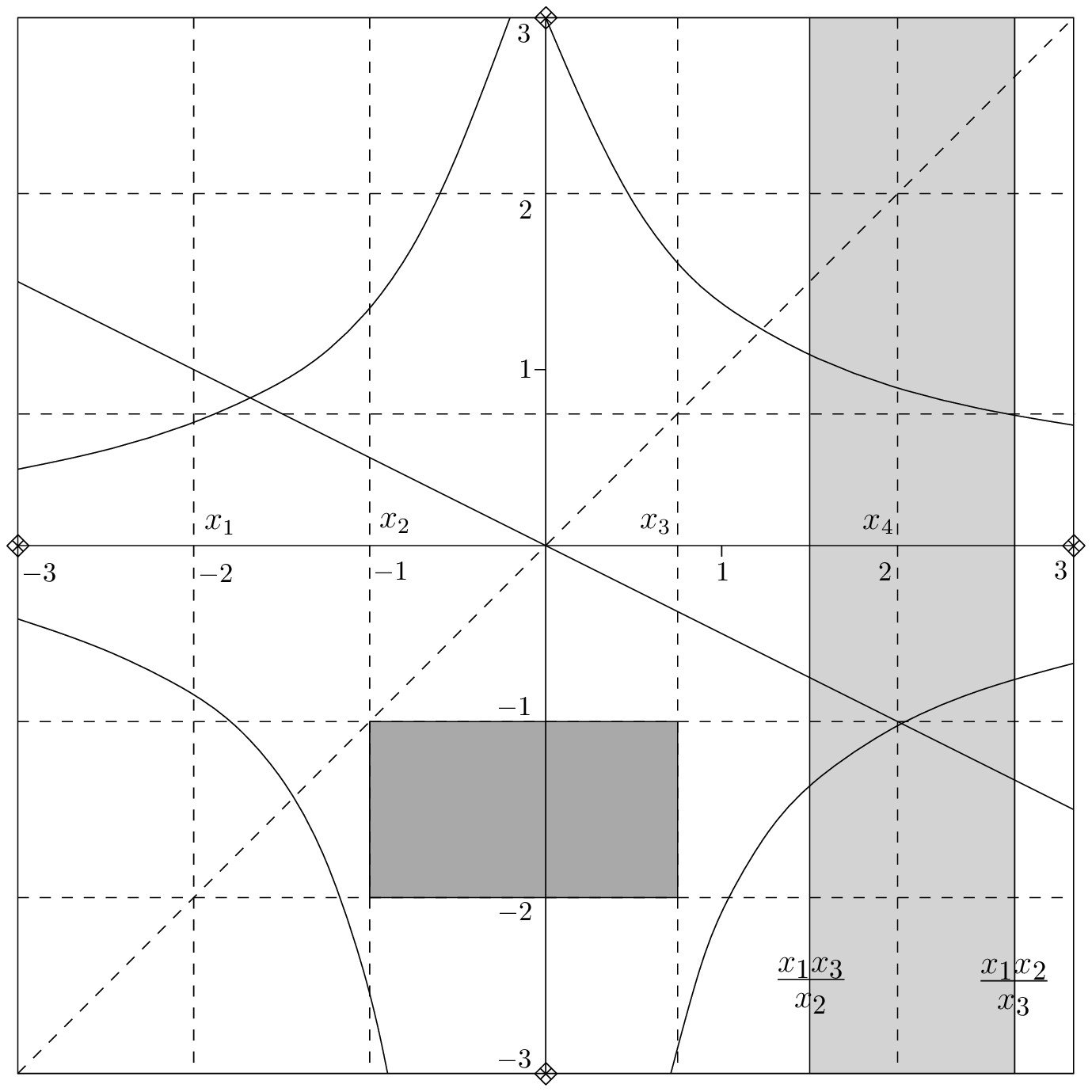}

~(a) \hskip2.95in (b)
\caption{The boxes for the cases (a) $x_3=-\frac{2}{3}$, $x_4=\frac{1}{2}$; and (b) $x_3=\frac{3}{4}$, $x_4=2$. The other free parameters have been set as $x_1=-2$ and $\nu=\frac{1}{2}$. The lighter-shaded strips represent the allowed ranges of $x_4$. The dashed diagonal line represents asymptotic infinity. The solid curves are locations of potential curvature singularities, as are the crossed diamonds at the four points ($x=0,y=\pm\infty$) and ($x=\pm\infty,y=0$). Note that the shaded box in each case touches asymptotic infinity at one corner while avoiding all the curvature singularities.}
\label{fig_box1}
\end{center}

\end{figure}

Hence we only need to consider a box touching the point $x=y=x_2$ at its upper-left corner. The analysis of this case can be carried out in a similar fashion. Since $(x=0,y=-\infty)$ is a curvature singularity, any box which does not contain this point has $x_1<x_2$, or can be transformed to such a form by applying the inversion-swapping symmetry. Furthermore, by requiring that $H$ is everywhere non-vanishing inside the box, one gets the constraint $x_2<x_3$. In fact, a detailed analysis shows that to avoid a possible curvature singularity inside the box, we need only to consider the box satisfying the following ranges of coordinates and parameters:
\be
\label{ranges_parameter1}
-\infty< x_1< y< x_2< x< x_3< -x_2\,,
\ee
with $x_4$ obeying
\be
\label{ranges_parameter2}
\Bigg\{\begin{array}{l}
        x_4\in (-\infty,\frac{x_1x_2}{x_3})\cup(\frac{x_2x_3}{x_1},\infty)\,,\qquad\,{\rm if} \,\,x_3< 0\,;\\
        x_4\in (\frac{x_1x_3}{x_2},\frac{x_1x_2}{x_3})\,,\hspace{2.97cm}\,{\rm if}\,\,x_3> 0\,.
      \end{array}
\ee
The special value $x_3=0$ can be recovered as a limit of either case. The two cases $x_3<0$ and $x_3>0$ are illustrated in the plots in Fig.~\ref{fig_box1}(a) and (b) respectively, for specific choices of parameters.

When $x_3< 0$, one can in fact further narrow the range of $x_4$ to the finite interval $(\frac{x_2x_3}{x_1},-x_2)$ by using a combination of the inversion-swapping and scaling symmetries, which maps a solution with any given $x_4=w$ to one with $x_4=\frac{x_2^2}{w}$ while preserving the value of $x_2$.

The conditions that we have derived are also sufficient to ensure that the signature of the metric is positive- or negative-definite in the region of interest. Since we have required that $X$ and $Y$ have opposite signs, it follows that $F$ defined in (\ref{metric_general}) has the same sign as $X$. If $k$ is positive, then the metric in (\ref{metric_general}) can be seen to have a positive- or negative-definite signature.

To summarise, we have the coordinate range $x_2< x< x_3$ and $x_1< y< x_2$, which can be visualised as a box in an $x$-$y$ plot.
 With the gauge choice (\ref{parameter_range2}), the solution is then characterised by five parameters: the position of the box determined by $x_{1,3}$, the fourth root $x_4$ of $X$, the line
\be
y=-\nu x\,,
\ee
with a slope between $-1$ and $1$, and an overall (positive) scale factor $k$. We have shown that these parameters have to satisfy (\ref{ranges_parameter1}) and (\ref{ranges_parameter2}), in order to ensure that the solution (\ref{metric_general}) has Euclidean signature and that it does not contain any curvature singularities.

\newsubsection{{ALF and non-self-dual properties}}

One key feature of the solution (\ref{metric_general}) is its ALF property (with AF as a special case). One can quickly calculate that when $x\rightarrow y$,
\be
\label{tau_norm}
g_{\tau\tau}\rightarrow \frac{1}{1-\nu^2}\,,
\ee
which is a finite constant, a characteristic of an ALF metric. This selects a particular Killing vector $\frac{\partial}{\partial \tau}$ (up to constant multiplication) for the solution, which has a finite norm and generates the Euclidean time flow at infinity. This property obviously indicates that, unlike say the Ricci-flat Pleba\'nski--Demia\'nski solution, there is no symmetry between the coordinates $\tau$ and $\phi$. To show the solution is ALF, one can define the coordinates ($r,\theta$) around the asymptotic region $x=y=x_2$ by
\be
x= x_2-\frac{x_2\sqrt{k(1-\nu^2)}}{r}\cos^2\frac{\theta}{2}\,,\quad y=x_2+\frac{x_2\sqrt{k(1-\nu^2)}}{r}\sin^2\frac{\theta}{2}\,,
\ee
and ($\tilde{\tau},\tilde{\phi}$) via the simultaneous substitutions
\be
\label{natural_killing_vectors}
\tau\rightarrow\sqrt{1-\nu^2}\,\tilde{\tau}+b\tilde{\phi}\,,\quad \phi\rightarrow-\frac{2\sqrt{k}x_2}{a_4(x_2-x_1)(x_2-x_3)(x_2-x_4)}\tilde{\phi}\,.
\ee
For large $r$, with $b$ chosen as
\be
b=\frac{\sqrt{k}[(1-\nu^2)(x_2^3+x_1x_3x_4)+2\nu x_2(x_1x_2+x_1x_3+x_1x_4+x_2x_3+x_2x_4+x_3x_4)]}{(x_1-x_2)(x_2-x_3)(x_2-x_4)}\,,
\ee
the metric approaches (\ref{definition_ALF}), with $(\tau,\phi)$ replaced by $(\tilde{\tau},\tilde{\phi})$ and
\be
n=\frac{\sqrt{k}[2\nu x_2(x_1x_2+x_2x_3+x_2x_4-x_1x_3-x_1x_4-x_3x_4)-(1+\nu^2)(x_1x_3x_4-x_2^3)]}{2\sqrt{1-\nu^2}(x_2-x_1)(x_2-x_3)(x_2-x_4)}\,,
\ee
or in an alternative form
\be
\label{total_nut_charge}
n=\frac{\sqrt{k}[2\nu(a_1x_2-a_3x_2^3)-(1-\nu)^2(a_0-a_4x_2^4)]}{2a_4\sqrt{1-\nu^2}x_2(x_2-x_1)(x_2-x_3)(x_2-x_4)}\,.
\ee
We refer to the Killing vectors associated to the newly defined coordinates ($\tilde{\tau},\tilde{\phi}$) as the ``natural Killing vectors'' of our solution, which can be seen to generate natural notions of (Euclidean) time translation and rotation respectively at infinity. In the limit $n\rightarrow \infty$, the solution should become ALE. Indeed, in the limits $\nu=\pm 1$, the solution reduces to the Ricci-flat Pleba\'nski--Demia\'nski and triple-collinearly-centered Gibbons--Hawking solutions respectively, both of which are ALE. The solution can thus be thought of as a one-parameter family of ALF metrics interpolating between the latter two metrics, indexed by $\nu$. On the other hand, the solution becomes AF in the limit $n=0$, and in particular it contains the AF instanton discovered in \cite{Chen:2011tc} as a special case. These important limits will be discussed in detail in Sec.~4.

The solution (\ref{metric_general}) has a Riemann tensor that is in general neither self-dual nor anti-self-dual. One self-dual limit is taken by setting $\nu=-1$, which, as just mentioned, gives the triple-collinearly-centered Gibbons--Hawking solution. There is another self-dual limit in the case when $X$ has two pairs of opposite roots, which gives the double-centered Taub-NUT solution. This limit will be discussed in detail in Appendix A.2.

\newsubsection{Rod structure}

The solution (\ref{metric_general}) possesses two apparent Killing vectors $\frac{\partial}{\partial \tau}$ and $\frac{\partial}{\partial \phi}$. It can thus be put in a canonical form, in so-called Weyl--Papapetrou coordinates given by
\be
\rho=\frac{\sqrt{-XY}}{(x-y)^2}\,,\quad z=\frac{2(a_0+a_2xy+a_4x^2y^2)+(x+y)(a_1+a_3xy)}{2(x-y)^2}\,,
\ee
where $z$ is determined up to a constant shift and a flip of sign. From the expression of $\rho$, one can see that the locations of the rods correspond to the roots (including possible infinite roots) of $X$ and $Y$. Turning points are then the meeting points of these rods: just like the Pleba\'nski--Demia\'nski solution, the above solution contains three turning points in general.

In the coordinates (\ref{metric_general}), the turning points are located at $(x=x_2,y=x_1)$, $(x=x_3,y=x_1)$ and $(x=x_3,y=x_2)$ respectively, or equivalently, along the $z$-axis with
\be
z_1=-\frac{a_4(x_1x_2+x_3x_4)}{2}\,,\quad z_2=-\frac{a_4(x_1x_3+x_2x_4)}{2}\,,\quad z_3=-\frac{a_4(x_1x_4+x_2x_3)}{2}\,,
\ee
respectively in Weyl--Papapetrou coordinates. The four rods have directions $\ell_i=K_i/\kappa_i$, with
\ba
K_1&=&\left(-\frac{-a_0+2\nu a_0+2\nu a_1x_2+\nu^2a_4x_2^4}{x_2^2},1\right),\quad \kappa_1=\frac{a_4(x_2-x_1)(x_2-x_3)(x_2-x_4)}{2\sqrt{k}x_2}\,,\cr
K_2&=&\left(-\frac{\nu^2a_0+2\nu a_3x_1^3+2\nu a_4x_1^4-a_4x_1^4}{x_1^2},1\right),\quad \kappa_2=\frac{a_4(x_1-x_2)(x_1-x_3)(x_1-x_4)}{2\sqrt{k}x_1}\,,\cr
K_3&=&\left(-\frac{-a_0+2\nu a_0+2\nu a_1x_3+\nu^2a_4x_3^4}{x_3^2},1\right),\quad \kappa_3=\frac{a_4(x_3-x_1)(x_3-x_2)(x_3-x_4)}{2\sqrt{k}x_3}\,,\cr
K_4&=&\left(-\frac{\nu^2a_0+2\nu a_3x_2^3+2\nu a_4x_2^4-a_4x_2^4}{x_2^2},1\right),\quad \kappa_4=\kappa_1\,.
\ea
Here, the direction $(\alpha,\beta)$ is defined by $\alpha\frac{\partial}{\partial\tau}+\beta\frac{\partial}{\partial\phi}$. Alternative expressions for $K_i[1]$ are
\ba
K_1[1]&=&a_4[x_1x_3x_4/x_2+2\nu (x_1x_3+x_1x_4+x_3x_4)-\nu^2 x_2^2]\,,\cr
K_2[1]&=&a_4[x_1^2+2\nu (x_1x_2+x_1x_3+x_1x_4)-\nu^2 x_2x_3x_4/x_1]\,,\cr
K_3[1]&=&a_4[x_1x_2x_4/x_3+2\nu (x_1x_2+x_1x_4+x_2x_4)-\nu^2 x_3^2]\,,\cr
K_4[1]&=&a_4[x_2^2+2\nu (x_1x_2+x_2x_3+x_2x_4)-\nu^2 x_1x_3x_4/x_2]\,.
\ea
These are also useful, since our analysis below will be based mainly on the factorised form of $X$ in (\ref{alt_param}), involving its four roots.

The above rod structure encodes much useful information about the solution. The relative directions of the four rods, and the ratio of the lengths of two finite rods are actually invariants of the solution. For example, the asymptotic NUT charge $n$ is related to the directions of the two asymptotic rods by $n=\lambda({K}_1[1]-{K}_{4}[1])$, where $\lambda$ is some proportionality constant. In the same spirit, one can define the NUT-charges carried by the individual turning points to be $n_i\equiv\lambda (K_i[1]-K_{i+1}[1])$, so that one has
\ba
\label{n123}
n_1&=&\frac{\sqrt{k}(x_1+\nu x_2)^2(x_1x_2-x_3x_4)}{2x_1\sqrt{1-\nu^2}(x_2-x_1)(x_2-x_3)(x_2-x_4)}\,,\cr
n_2&=&\frac{x_2\sqrt{k}(x_1+\nu x_3)^2(x_2x_4-x_1x_3)}{2x_1x_3\sqrt{1-\nu^2}(x_2-x_1)(x_2-x_3)(x_2-x_4)}\,,\cr
n_3&=&\frac{\sqrt{k}(x_2+\nu x_3)^2(x_2x_3-x_1x_4)}{2x_3\sqrt{1-\nu^2}(x_2-x_1)(x_2-x_3)(x_2-x_4)}\,.
\ea
Note that the asymptotic NUT charge is then the sum of the individual NUT charges: $n=n_1+n_2+n_3$.

The rod structure will be extensively used in the subsequent study. In particular, many special cases of the general solution were identified by first studying the behaviour of the rod structure. In the rest of this subsection, we will briefly discuss the special case when two adjacent rods are joined up.

Two adjacent rods, say the $i$-th and $i+1$-th rods, can be joined up by setting $n_i=0$. In this case, the turning point at which they meet is effectively eliminated, and one obtains a solution whose rod structure has only two turning points. For example, we can impose
\be
\label{n3=0}
n_3=0\,,
\ee
to join up the third and fourth rods, thus eliminating the third turning point from the rod structure. An obvious solution to this condition is
\be
\label{join_rods34}
x_1x_4=x_2x_3\,.
\ee
The resulting solution is the Kerr-NUT solution. In this limit, the box touches a solution curve of $H$ at one of its corners. We have mentioned that the solution curves are locations of curvature singularities for general parameters. But in the present case, it can be explicitly checked that at this corner, a zero factor emerges from the numerator of the Kretschmann invariant, which cancels the zero factor of $H$. This gives rise to a finite Kretschmann invariant. Another solution to the condition (\ref{n3=0}) is
\be
x_3\rightarrow -x_2\,,\quad x_4\rightarrow -x_1\,,\quad \nu\rightarrow 1\,,
\ee
which results in the double-centered Taub-NUT solution. This limit can also be understood as a box with one corner touching a solution curve of $H$, now corresponding to its factor $\nu x+y$. The details of how the Kerr-NUT and double-centered Taub-NUT solutions are recovered in these limits can be found in Appendix A.1 and A.2, respectively.

The joining-up of the first and second rods, or of the second and third rods, can be similarly obtained. We can see that the (finite) lower and upper bounds previously identified for the parameter $x_4$ in (\ref{ranges_parameter2}) correspond exactly to the joining up of different pairs of adjacent rods to obtain the Kerr-NUT solution.

\newsection{Various limits}

We have noted that the solution (\ref{metric_general}) is ALF, and that it contains two ALE limits: the Pleba\'nski--Demia\'nski solution and the triple-collinearly-centered Gibbons--Hawking solution. In this section, we show in detail how these two limits are recovered. The AF limit of (\ref{metric_general}) is also discussed, and we show how the new AF gravitational instanton can be recovered from it. Finally, a new special case of (\ref{metric_general}) is discussed: an ALF generalisation of the C-metric, which can also be called the NUT-charged C-metric. Note that all the limits discussed here have three turning points in their rod structure; limits with two or one turning points will be discussed in Appendix A.

\newsubsection{Pleba\'nski--Demia\'nski solution}

Recall from (\ref{tau_norm}) that the norm of $\frac{\partial}{\partial\tau}$ at infinity becomes unbounded when $\nu\rightarrow \pm1$. This means that the metric becomes ALE. Here, we will show that by directly setting
\be
\nu= 1\,,
\ee
we recover the Pleba\'nski--Demia\'nski metric from (\ref{metric_general}). We first perform the M\"obius transformation
\be
x=\frac{p+1}{p-1}\,,\quad y=\frac{q+1}{q-1}\,,
\ee
and redefine the parameters
\be
a_{0,4}={\gamma\over2}-{\epsilon\over4}\mp{m+n\over2}\,,\quad a_{1,3}=\mp (m-n)\,,\quad a_2={\epsilon\over2}+3\gamma\,,\quad k=\frac{1}{m-n}\,.
\ee
Note that this redefinition of six parameters in terms of only four is consistent with the above-mentioned fact that two of the original parameters are redundant. Then after a linear transformation of the coordinates $\tau$ and $\phi$ via the simultaneous substitutions
\be
\tau\rightarrow  \frac{(4m-4n+6\gamma+\epsilon)\tau+(4m-4n-6\gamma-\epsilon)\phi}{4\sqrt{m-n}}\,,\quad \phi\rightarrow\frac{\tau-\phi}{2\sqrt{m-n}}\,,
\ee
the metric is brought precisely to the Ricci-flat class of the (Euclideanlised) Pleba\'nski--Demia\'nski metric \cite{Plebanski:1976gy}
\ba
\label{PD}
\dif s^2&=&\frac{1}{(p-q)^2}\Big[\frac{1-p^2q^2}{P}\dif p^2+\frac{P}{1-p^2q^2}(\dif \phi-q^2\dif \tau)^2\cr
&&\hspace{0.56in}-\frac{1-p^2q^2}{Q}\dif q^2-\frac{Q}{1-p^2q^2}(\dif \tau-p^2\dif \phi)^2\Big]\,,\cr
P&=&\gamma+2np-\epsilon p^2+2mp^3+\gamma p^4,\quad Q=\gamma+2nq-\epsilon q^2+2mq^3+\gamma q^4.
\ea

\newsubsection{Triple-collinearly-centered Gibbons--Hawking solution}

The other ALE limit is taken by directly setting
\be
\nu= -1\,,
\ee
and one recovers the triple-collinearly-centered Gibbons--Hawking solution. To cast the resulting solution in a more familiar form, we first define the coordinates $(r,\theta)$ by
\be
\label{rtheta}
r\sin\theta=\frac{\sqrt{-XY}}{(x-y)^2}\,,\quad r\cos\theta=\frac{2(a_0+a_2xy+a_4x^2y^2)+(x+y)(a_1+a_3xy)}{2(x-y)^2}\,,
\ee
the parameters $d_{1,2,3}$ in terms of the roots of $X$ as
\ba
\label{d123}
d_1&=&-\frac{a_4(x_1x_2+x_3x_4)}{2}\,,\quad d_2=-\frac{a_4(x_1x_3+x_2x_4)}{2}\,,\quad d_3=-\frac{a_4(x_1x_4+x_2x_3)}{2}\,,
\ea
and $n_{1,2,3}$ as
\ba
\label{NUT_triple_BH}
n_1&=&\frac{k(x_1x_2-x_3x_4)}{a_4(x_1-x_3)(x_1-x_4)(x_2-x_3)(x_2-x_4)}\,,\cr
n_2&=&\frac{k(x_1x_3-x_2x_4)}{a_4(x_1-x_2)(x_1-x_4)(x_2-x_3)(x_3-x_4)}\,,\cr
n_3&=&\frac{k(x_2x_3-x_1x_4)}{a_4(x_1-x_2)(x_1-x_3)(x_2-x_4)(x_3-x_4)}\,.
\ea
After the simultaneous substitutions
\be
\tau\rightarrow (-a_2\tau+4k\phi)/\sqrt{4k}\,,\quad \phi\rightarrow \tau/\sqrt{4k}\,,
\ee
the metric is brought to the form
\ba
{\dif s}^{2}&=& V^{-1} \left( {\dif\tau}+A \right) ^{2}+V
 ( {\dif r}^{2}+{r}^{2} {{\dif\theta}}^{2}+ r^2 \sin ^{2}
 \theta\, {{\dif\phi}}^{2}  )\,,\cr
 V&=&\sum_{i=1}^3 \frac{2n_i}{r_i}\,,\quad A= \sum_{i=1}^3 \frac{2 n_i (r \cos \theta-d_i)}{r_i}\,\dif\phi\,,\quad r_i=\sqrt {{r}^{2}+{{d_i}}^{2}-2 {d_i} r \cos \theta}\,,
\ea
which is recognised as the triple-collinearly-centered Gibbons--Hawking solution. Note that there is a translational symmetry by adding an arbitrary constant to the right-hand sides of $r\cos\theta$ in (\ref{rtheta}) and $d_{1,2,3}$ in (\ref{d123}). The charges $n_{1,2,3}$ in (\ref{NUT_triple_BH}) always have the same sign in the ranges (\ref{ranges_parameter1}) and (\ref{ranges_parameter2}) that we considered, as required by the absence of curvature singularities in the solution.

\newsubsection{AF limit and the new AF instanton}

This limit is taken by setting the total NUT-charge (\ref{total_nut_charge}) to zero, so now the metric (\ref{metric_general}) becomes AF. The solution to the condition $n=0$ and the gauge choice $x_2=-1$ can be written as follows:
\be
a_1+a_3=a_0+a_2+a_4\,,\quad
a_1-a_3=-\frac{(1-\nu)^2(a_0-a_4)}{2\nu}\,.
\ee

Within this four-parameter AF subclass is the completely regular gravitational instanton recently discovered by the present authors \cite{Chen:2011tc} and christened the ``new AF instanton''. It can be obtained as a special case by further imposing
\be
\ell_1=\ell_4=\pm\ell_2\pm\ell_3\,.
\ee
These conditions ensure that the metric is defined on an underlying manifold with a globally well-defined $U(1)\times U(1)$ isometry, and is free of conical and orbifold singularities. A solution to these conditions can be written (for general $x_2$) in the following parametric forms:
\be
\nu=-{2\xi^2},\quad x_1=-\frac{\xi(1-2\xi+2\xi^2)x_2}{1-2\xi}\,,\quad x_3=\frac{(1-2\xi+2\xi^2)x_2}{4\xi^2(1-\xi)}\,,\quad x_4=0\,,
\ee
or
\be
\nu=-{2\xi^2},\quad x_1=\frac{4\xi^2(1-\xi)x_2}{1-2\xi+2\xi^2}\,,\quad x_3=-\frac{(1-2\xi)x_2}{\xi(1-2\xi+2\xi^2)}\,,\quad x_4=\infty\,,
\ee
which are related to each other by the inversion-swapping symmetry.

In the latter parameterisation with an appropriate gauge choice, the new AF instanton is then given by the metric (\ref{metric_general}) with the parameter $\nu$ and the structure function $X$ defined as
\be
\nu=-2\xi^2,\quad X=(x+4\xi^3-4\xi^4)(x+\xi-2\xi^2+2\xi^3)(x-1+2\xi)\,.
\ee
The parameters $a_{0,\dots,4}$ are encoded in the polynomial $X$ as its coefficients; in particular, we simply have $a_4=0$ and a gauge-fixed $a_3=1$. The solution is then determined by the parameters $k$ and $\xi$. This gives an alternative but simpler form of the new AF instanton. Now the awkward square-root terms in the latter are eliminated and its analysis based on this parameterisation will be much easier. To cast it in exactly the same form used in \cite{Chen:2011tc}, one needs to perform the M\"obius transformation:
\be
x\rightarrow-\frac{\xi[(1-2\xi+2\xi^2)^2x-(1-2\xi^2)(1-4\xi+2\xi^2)]}{(1-2\xi+2\xi^2)x+1-2\xi^2}\,,
\ee
with a similar equation for $y$, followed by the simultaneous substitutions
\ba
\tau&\rightarrow&\sqrt{1-4\xi^4}\,\psi-\frac{16k^2\xi^3(1-4\xi+8\xi^2-12\xi^3+16\xi^4-8\xi^5)}{\sqrt{1-4\xi^4}(1-2\xi^2)(1-2\xi+2\xi^2)^2}\phi\,,\cr
\phi&\rightarrow&\frac{4k^2}{\sqrt{1-4\xi^4}(1-2\xi^2)(1-2\xi+2\xi^2)}\phi\,,\cr
k&\rightarrow &\frac{4k^4(1-2\xi^2)(1-\xi)^2(1-2\xi)^2}{(1+2\xi^2)(1-2\xi+2\xi^2)^4}\,,
\ea
and finally introduce the new parameters $\lambda$ and $\gamma$ defined as
\be
\lambda=\frac{1-2\xi^2}{1-2\xi+2\xi^2}\,,\quad \gamma=-\frac{1-4\xi+2\xi^2}{1-2\xi+2\xi^2}\,.
\ee

\newsubsection{NUT-charged C-metric}

The NUT-charged C-metric is obtained by requiring that the second rod is parallel to the fourth rod, which is achieved by setting
\be
a_4=0~\,{\rm or}~\,x_4=\pm \infty\,,\quad \nu=0\,.
\ee
The metric then becomes
\ba
\label{metric_NUT_C-metric}
\dif s^2&=&\frac{\left(F\dif \tau+a_0Y\dif \phi\right)^2}{(x-y)HF}+\frac{kH}{(x-y)^3}\left(\frac{\dif x^2}{X}-\frac{\dif y^2}{Y}-\frac{XY}{kF}\,\dif \phi^2\right),\cr
H&=&y(-2a_0-a_1y+a_3xy^2)\,,\quad F=y^2X-x^2Y\,,\cr
X&=&a_0+a_1x+a_2x^2+a_3x^3,\quad Y=a_0+a_1y+a_2y^2+a_3y^3.
\ea
We make the gauge choice $a_3=1$ and use the roots of $X=(x-x_1)(x-x_2)(x-x_3)$ as the parameters. Requiring $H$ to be everywhere non-vanishing in the box gives the following ranges of the parameters and coordinates:
\be
-\infty< x_1< y< x_2< x< x_3< 0\,,
\ee
which is consistent with (\ref{ranges_parameter1}) and (\ref{ranges_parameter2}) from our general analysis. The reader is reminded that we actually have $x_4=-\infty$, which provides the lower bound for $x_1$ and implies the upper bound for $x_3$.

The rod structure of this solution can be brought to a form very close to that of the C-metric by defining
\be
\tau=\tau',\quad \phi=-\frac{2x_2\sqrt{k}}{(x_1-x_2)(x_2-x_3)}{\phi}'.
\ee
In these coordinates, the corresponding rod directions are
\ba
\ell_1=(4n,1)\,,\quad \ell_2=\frac{1}{{\kappa}'_2}(0,1)\,,\quad \ell_3=\frac{1}{{\kappa}'_3}(4n',1)\,,\quad \ell_4=(0,1)\,,
\ea
where $n$ and $\kappa'_2$ are given by
\be
n=\frac{\sqrt{k}x_1x_3}{2(x_1-x_2)(x_2-x_3)}\,,\quad {\kappa}'_2=\frac{x_2(x_1-x_3)}{x_1(x_2-x_3)}\,,
\ee
and $n'$ and $\kappa'_3$ are given by
\be
\label{n'}
n'=\frac{x_2^2}{x_3^2}n\,,\quad\frac{1}{\kappa'_2}+\frac{1}{{\kappa}'_3}=1\,.
\ee
We note that the lengths of the two finite rods are respectively
\be
{z}'_{21}=\frac{\sqrt{k}x_2}{x_1-x_2}\,,\quad {z}'_{32}=\frac{\sqrt{k}x_2}{x_2-x_3}\,.
\ee
This rod structure is illustrated in Fig.~2. The first and third rods are the analogues of the horizon rods of the C-metric, while the second and fourth rods are the usual axis rods.

\begin{figure}[!t]

\begin{center}
  \includegraphics{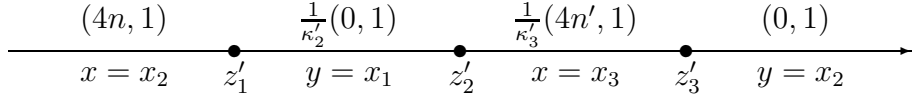}
 \end{center}
\caption{The rod structure of the NUT-charged C-metric, with asymptotic NUT charge $n$.}

\end{figure}

Like the original solution, the metric (\ref{metric_NUT_C-metric}) is ALF with an asymptotic NUT charge $n$. An ALE limit is obtained when $n\rightarrow \infty$, in which case we recover the C-metric. To take this limit, we need to set
\be
x_{1,3}\rightarrow x_2\,,
\ee
while fixing the other parameters. The box in this case thus becomes a point at $(x=x_2,y=x_2)$. More specifically, we first write the parameters in the following form:
\be
x_1=x_2(1+(1-c)\epsilon)\,,\quad x_3=x_2(1-2c\epsilon)\,,\quad k= \epsilon^3c^2l^2,
\ee
and define new coordinates by performing the following substitutions:
\ba
x&\rightarrow& x_2[1-c(1+x)\epsilon]\,,\quad y\rightarrow x_2[1-c(1+y)\epsilon]\,,\cr
\tau&\rightarrow& \frac{2x_1x_3\sqrt{k}(1-c)}{(x_1-x_2)(x_2-x_3)}\,\tau\,,\quad \phi\rightarrow -\frac{2x_2\sqrt{k}(1-c)}{(x_1-x_2)(x_2-x_3)}\,(\tau+\phi)\,.
\ea
After taking the limit $\epsilon\rightarrow 0$, the metric becomes the familiar form of the C-metric:
\be
\dif s^2=\frac{l^2}{(x-y)^2}\left[\frac{\dif x^2}{G(x)}+G(x)\dif \phi^2-\frac{\dif y^2}{G(y)}-G(y)\dif \tau^2\right],\quad G(x)=(1-x^2)(1+cx)\,.
\ee

On the other hand, the AF limit of (\ref{metric_NUT_C-metric}) is taken by requiring
\be
n=0 \quad \Rightarrow\quad x_3=0\quad {\rm or} \quad a_0=0\,.
\ee
In this limit, the first rod joins up with the second, or in other words, the first turning point vanishes. The resulting rod structure has two turning points, and turns out to be just the Schwarzschild solution. The mapping to this metric is a special case of the transformation described in Appendix A.3.

The solution (\ref{metric_NUT_C-metric}) admits several other, more subtle limits. In the limit $x_1\rightarrow x_2$ and $k\rightarrow 0$ while keeping $\frac{(x_1-x_2)^2}{k}$ finite, one can recover the self-dual Taub-NUT solution. In this case, the second and third turning points merge with each other and together they disappear from the rod structure. The details of how this limit is taken can be found in Appendix A.4. The self-dual Taub-NUT solution can also be obtained in the limit in which the second and third turning points are pushed to infinity. These limits show that the asymptotic NUT charge can be thought of as being carried by the first turning point.

Finally, in the limit $x_1=-\infty$ or $a_3=0$, the second rod shrinks down to zero length, and the first and second turning points merge with each other. In this case, one recovers the Kerr-NUT solution with NUT-charge parameter equal to the rotational parameter, i.e., (\ref{Euclidean Kerr-bolt}) with $n=a$.

\newsection{Alternative interpretation as two touching Kerr-NUTs}

In this section, we show how the general solution (\ref{metric_general}) can be regarded as a system consisting of two touching Kerr-NUTs. We also analyse in detail, the special case consisting of a Schwarzschild in superposition with a self-dual Taub-NUT.

\newsubsection{Kerr-NUT touching Kerr-NUT}

The general solution (\ref{metric_general}) has an alternative interpretation as a system consisting of two touching Kerr-NUTs: the south pole of one Kerr-NUT touches the north pole of the other. This is consistent with our ISM construction. Recall that in our construction, one possible seed solution is the double Schwarzschild solution, with an inner axis separating the two black holes; the inner axis is then joined up with one of the black-hole horizons by appropriately choosing the corresponding BZ parameter. The inner axis disappears from the solution, and in this sense the resulting two Kerr-NUTs are touching each other, sharing one single point as their common pole. In what follows, we shall present the rod structure of the solution in a form that favours this interpretation, and show how one can remove one Kerr-NUT from the solution in certain limits.

In the natural Killing coordinates, the rod structure is
\be
\tilde{\ell}_{1}=(2n,1)\,,\quad
\tilde{\ell}_{2}=\frac{1}{\kappa_{\rm I}}(1,\Omega_{\rm I})\,,\quad
\tilde{\ell}_{3}=\frac{1}{\kappa_{\rm II}}(1,\Omega_{\rm II})\,,\quad
\tilde{\ell}_{4}=(-2n,1)\,,
\ee
where the angular velocities $\Omega_{\rm I,II}$ are given by
\ba
\Omega_{\rm I}&=&-\frac{\sqrt{1-\nu^2}x_1x_3(x_2-x_1)(x_2-x_3)(x_2-x_4)}{\sqrt{k}(-L_1+L_2+L_3)}\,,\cr
\Omega_{\rm II}&=&-\frac{\sqrt{1-\nu^2}x_1x_3(x_2-x_1)(x_2-x_3)(x_2-x_4)}{\sqrt{k}(-L_1-L_2+L_3)}\,,
\ea
and the surface gravities $\kappa_{\rm I,II}$ are given by
\ba
\kappa_{\rm I}&=&\frac{\sqrt{1-\nu^2}x_2x_3(x_1-x_2)(x_1-x_3)(x_1-x_4)}{\sqrt{k}(-L_1+L_2+L_3)}\,,\cr
\kappa_{\rm II}&=&\frac{\sqrt{1-\nu^2}x_1x_2(x_3-x_1)(x_3-x_2)(x_3-x_4)}{\sqrt{k}(-L_1-L_2+L_3)}\,.
\ea
Here, $L_{1,2,3}$ are the three functions defined as
\ba
L_1&=&x_1(x_2x_3-x_1x_4)(x_2+\nu x_3)^2,\quad L_2=x_2(x_1x_3-x_2x_4)(x_1+\nu x_3)^2,\cr L_3&=&x_3(x_1x_2-x_3x_4)(x_1+\nu x_2)^2.
\ea
We note that the lengths of the rods are
\be
\tilde{z}_{21}=\frac{\sqrt{k-k\nu^2}x_2(x_1-x_4)}{(x_1-x_2)(x_2-x_4)}\,,\quad \tilde{z}_{32}=\frac{\sqrt{k-k\nu^2}x_2(x_3-x_4)}{(x_2-x_3)(x_2-x_4)}\,.
\ee
This rod structure is illustrated in Fig.~3. For comparison, we recall that the rod structure of a single Kerr-NUT \cite{Chen:2010zu} has three rods, with directions $(2n,1)$, ${1\over\kappa}(1,\Omega)$ and $(-2n,1)$ in suitable coordinates, where $n$ is the NUT charge, and $\kappa$ and $\Omega$ are the surface gravity and angular velocity of the horizon respectively. It is then clear that the rod structure considered here consists of two Kerr-NUTs, whose horizons touch at the poles. $\kappa_{\rm I,II}$ and $\Omega_{\rm I,II}$ are the surface gravities and angular velocities of the two horizons respectively.

\begin{figure}[!t]

\begin{center}
  \includegraphics{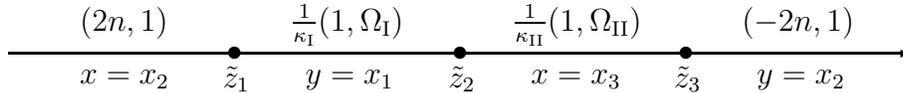}
\end{center}
\caption{The rod structure of the general solution, regarded as a touching double-Kerr-NUT system.}

\end{figure}

One can remove a Kerr-NUT from the solution by zooming in to one of the Kerr-NUTs and at the same time sending the other pole of the second Kerr-NUT to infinity. Alternatively, one can shrink the size of the second Kerr-NUT to zero. We will give a more detailed discussion of both of these possibilities in Appendix A.1. The explicit mapping of the first limit to the Kerr-NUT solution is given therein. Recall that in addition to these two possibilities, we have a third possibility to recover Kerr-NUT by joining up two rods as discussed in Sec.~3.3, corresponding to the lower and upper bounds of $x_4$.

In the rest of this subsection, we will briefly discuss a few special cases of this solution. First, consider the case
\be
x_1=\infty\,,~\, {\rm i.e.,}~\, a_4=0\quad \Rightarrow\quad  \Omega_{\rm I}=0\,,
\ee
in which the first Kerr-NUT becomes static.\footnote{Note that there is a well-defined notion of being ``static'', given by the requirement that the rod has a direction proportional to $\frac{\partial}{\partial \tau}$.} This can be interpreted as a configuration in which a (non-self-dual) Taub-NUT touches a Kerr-NUT. On the other hand, the case
\be
x_3=0\,,~\, {\rm i.e.,}~\, a_0=0\quad \Rightarrow \quad \Omega_{\rm II}=0\,,
\ee
is essentially the same configuration, with the locations of these two objects swapped. This is of course expected, since they are related by the inversion-swapping symmetry.

The case with both Kerr-NUTs static:
\be
x_1=\infty\,,~x_3=0\,,~\, {\rm i.e.,}~\, a_{0,4}=0\quad \Rightarrow\quad  \Omega_{\rm I,II}=0\,,
\ee
can then be interpreted as a configuration in which a Taub-NUT touches another Taub-NUT, which is of course a single larger Taub-NUT. We will discuss this limit in detail in Appendix A.3, and show how the familiar form of the Taub-NUT solution can be recovered. By further setting $\nu=0$, one obtains the Schwarzschild solution, which can be viewed as resulting from a Schwarzschild touching another Schwarzschild.

\newsubsection{Schwarzschild in superposition with a self-dual Taub-NUT}

The case describing a Schwarzschild in superposition with a self-dual Taub-NUT is obtained by requiring that the second rod is static and that the first rod is parallel to the third rod. The solution to these conditions is
\be
\label{temp3}
x_1=-\infty\,,\quad x_4=-\frac{2\nu x_2x_3}{x_2+x_3}\,.
\ee
An equivalent solution can be obtained by applying the inversion-swapping symmetry, in which the third rod becomes static and the second rod becomes parallel to the fourth rod. To be concrete, we will focus on the former case in the rest of this subsection.

Choosing the gauge $a_3=-x_2-x_3$, $x_2=-1$, and defining $x_3=\mu$, the metric is then given by
\ba
\label{metric_Sch-TN}
\dif s^2&=&\frac{\left(F\dif \tau+G\dif \phi\right)^2}{(x-y)HF}+\frac{kH}{(x-y)^3}\left(\frac{\dif x^2}{X}-\frac{\dif y^2}{Y}-\frac{XY}{kF}\,\dif \phi^2\right),\cr
H&=&(\nu x+ y)[(1-\mu)(\nu x-y)(\mu(2\nu-1)-xy)+4\mu^2\nu(1-\nu)]\,,\cr
G&=&2\nu[(y^3(1-\mu)-\mu^2\nu^2)X-\mu(2\nu-1)(x-\mu x-\mu)Y]\,,\quad F=y^2X-x^2Y\,,\cr
X&=&(x+1)(x-\mu)(x-\mu x+2\mu\nu)\,,\quad Y=(y+1)(y-\mu)(y-\mu y+2\mu\nu)\,.
\ea
Recall that $(x=0,y=-\infty)$ is a curvature singularity, so one needs to impose $\mu< 0$. The requirement that $H$ remains non-vanishing inside the box gives $\nu>0$. Then the ranges of the parameters and coordinates are given by
\be
\label{metric_Sch-TN_range}
-1< \mu< 0< \nu< 1\,,\quad -\infty< y< -1< x< \mu\,.
\ee
These ranges are consistent with (\ref{ranges_parameter1}) and (\ref{ranges_parameter2}) from our general analysis. The restriction on the range of the parameter $\nu$ originates from its relation to $x_4$ in (\ref{temp3}).

The rod structure of this solution can be brought to a nicer form by defining
\be
\tau=\sqrt{1-\nu^2}\,{\tau}'+\frac{4\mu\nu(2\nu-1)\sqrt{k}}{(1+\mu)(1-\mu-2\mu\nu)}{\phi}',\quad \phi=\frac{2\sqrt{k}}{(1+\mu)(1-\mu-2\mu\nu)}{\phi}'.
\ee
In these coordinates, the four rods are located at $x=-1$, $y=-\infty$, $x=\mu$ and $y=-1$ respectively. The corresponding rod directions are
\ba
\ell_1=(0,1)\,,\quad \ell_2=\frac{1}{{\kappa}'_2}(1,0)\,,\quad \ell_3=\frac{1}{{\kappa}'_3}(0,1)\,,\quad \ell_4=(-4n,1)\,,
\ea
where
\be
\label{temp1}
{\kappa}'_2=\frac{\sqrt{1-\nu^2}}{2\sqrt{k}}\,,\quad{\kappa}'_3=\frac{1-\mu+2\nu}{1-\mu-2\mu\nu}\,,\quad n=-\frac{\sqrt{k}\nu(1-\mu\nu)^2}{\sqrt{1-\nu^2}(1+\mu)(1-\mu-2\mu\nu)}\,.
\ee
We note that the lengths of the two finite rods are respectively
\be
\label{temp2}
{z}'_{21}=\frac{\sqrt{k(1-\nu^2)}(1-\mu)}{1-\mu-2\mu\nu}\,,\quad {z}'_{32}=\frac{\sqrt{k(1-\nu^2)}\mu(\mu-1-2\nu)}{(1+\mu)(1-\mu-2\mu\nu)}\,.
\ee
This rod structure is illustrated in Fig.~4. We recognise a Schwarzschild-like structure in the first three rods, while the third turning point can be regarded as an isolated self-dual Taub-NUT. The distance between the Schwarzschild horizon and the self-dual Taub-NUT is given by ${z}'_{32}$. In this picture, the asymptotic NUT charge $n$ is carried entirely by the self-dual Taub-NUT.

\begin{figure}[!t]

\begin{center}
  \includegraphics{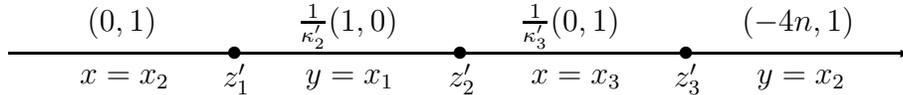}
\end{center}
\caption{The rod structure of a Schwarzschild in superposition with a self-dual Taub-NUT. In the parameterisation used here, one has $x_1=-\infty$, $x_2=-1$ and $x_3=\mu$.}

\end{figure}

Note that in the ranges of interest (\ref{metric_Sch-TN_range}), the quantities ${\kappa}'_2$, ${\kappa}'_3$, ${z}'_{21}$ and ${z}'_{32}$ remain positive. The NUT-charge $n$ is negative, and ranges from $0$ down to $-\infty$ when $\nu$ ranges from $0$ to $1$. We can thus interpret $\nu$ as the asymptotic NUT-charge parameter: $\nu=0$ corresponds to a zero value of the NUT charge and gives the Schwarzschild solution, while $\nu=1$ corresponds to an infinite (negative) value of the NUT charge and gives an ALE solution. $k$ simply sets the scale of the solution and can thus be taken as the mass parameter. $\mu$ determines the length of the third rod, and so can be interpreted as the separation parameter: $\mu=-1$ corresponds to the infinite-separation limit ${z}'_{32}=\infty$, and gives either the self-dual Taub-NUT or the Schwarzschild solution; while $\mu=0$ corresponds to the zero-separation limit ${z}'_{32}=0$, in which the self-dual Taub-NUT merges with the Schwarzschild to give a non-self-dual Taub-NUT. We do not present the details here, apart from mentioning that some of these limits are special cases of those considered in Appendix A.

One could wonder what happens when $-1\leq \nu< 0$, and in particular when $\nu=-1$. It turns out that these possibilities do not give rise to well-behaved solutions, since a curvature singularity appears inside the box. In the case $\nu=-1$, the above solution reduces to a triple-collinearly-centered Gibbons--Hawking solution with two nuts possessing opposite NUT charges. The latter is known to be singular, since it corresponds to at least one nut possessing a negative ``magnetic mass''.

\newsection{Interpretation in Kaluza--Klein theory}

Given an ALF solution, there is a well-known procedure by which it can be turned into a solution of four-dimensional Kaluza--Klein theory. One first adds a flat time direction to the ALF solution to obtain a solution of five-dimensional vacuum Einstein gravity with Lorentzian signature. Upon dimensional reduction along the compact direction $\frac{\partial}{\partial\tau}$ (or $\frac{\partial}{\partial\tilde\tau}$), one then obtains an asymptotically flat solution of Kaluza--Klein theory with Lorentzian signature.

When this procedure is applied to the Taub-NUT solution, one obtains a magnetic monopole solution in Kaluza--Klein theory \cite{Gross:1983hb,Sorkin:1983ns}. In this process, the NUT charge has turned into a magnetic charge associated with the Kaluza--Klein gauge field. When applied to the $n$-centered Taub-NUT solution, a system of $n$ magnetic monopoles results. Each turning point in the rod structure of the original solution has turned into a source for a monopole.

This procedure can also be applied to AF solutions with vanishing asymptotic NUT charge. The resulting Kaluza--Klein solution will then have zero total magnetic charge, although they would still in general describe a system of monopoles. For example, when this procedure is applied to the Kerr solution, one obtains a magnetic dipole \cite{Gross:1983hb}.

It is clear then that the solution (\ref{metric_general}) can be turned into a Kaluza--Klein solution describing a system of three collinearly centered magnetic monopoles. Of the five parameters in the original solution, three of them characterise the independent charges of the monopoles, and the remaining two their relative positions. The charges of these monopoles, as well as other properties of this system, can be read off from the rod structure of the original solution. In the natural Killing coordinates ($\tilde{\tau},\tilde{\phi}$), the four rods have directions $\ell_i=\tilde K_i/\tilde\kappa_i$, with
\ba
\tilde{K}_1&=&\left(2n_1+2n_2+2n_3,1\right),\quad\,\,\,\,\, \tilde{K}_2=\left(-2n_1+2n_2+2n_3,1\right),\cr
\tilde{K}_3&=&\left(-2n_1-2n_2+2n_3,1\right),\quad \tilde{K}_4=\left(-2n_1-2n_2-2n_3,1\right),
\ea
where $n_{1,2,3}$ are the NUT charges of the turning points, given by (\ref{n123}). Recall that the asymptotic NUT charge is $n=n_1+n_2+n_3$. The surface gravities in these coordinates are given by
\be
\tilde{\kappa}_{1,4}=1\,,\quad \tilde{\kappa}_2=\frac{x_2(x_1-x_3)(x_1-x_4)}{x_1(x_2-x_3)(x_2-x_4)}\,,\quad \tilde{\kappa}_3=\frac{x_2(x_3-x_1)(x_3-x_4)}{x_3(x_2-x_1)(x_2-x_4)}\,.
\ee

Now, $n_{1,2,3}$ can be identified with the charges of the three monopoles. The identity $\tilde{\kappa}_{1,4}=1$ indicates that the space-time described by the solution in Kaluza--Klein theory is asymptotically flat if $\tilde{\phi}$ has standard periodicity $2\pi$. Conical singularities then exist along the inner axes if $|\tilde{\kappa}_i|\neq1$, for $i=2,3$. For the ranges we consider, we actually always have $|\tilde{\kappa}_{2,3}|\geq 1$, with equality holding only in certain degenerate limits. So in general, one cannot obtain a balanced triple-monopole system in Kaluza--Klein theory from (\ref{metric_general}). Conical singularities necessarily exist along the two axes joining the monopoles. This is in contrast to the triple-centered Taub-NUT solution, which describes a balanced triple-monopole system.

Finally, we remark that in the AF limit $n=0$, the three monopoles will carry a zero total charge. A special case of this system was previously considered in \cite{Chen:2011tc}.

\newsection{Future work}

To summarise, we have presented a five-parameter class of Ricci-flat solutions in four dimensions with Euclidean signature. This solution can be regarded as a generalisation of the Ricci-flat Pleba\'nski--Demia\'nski solution with the inclusion of an asymptotic NUT charge; alternatively, it can be regarded as a system consisting of two touching Kerr-NUTs. Various properties and limits of the solution were studied.

The reader should bear in mind that we have entirely focused on the solution as a Euclidean solution. The most prominent problem related to this solution is whether it admits a Lorentzian section and how to find it if it exists. Such a section would be very interesting, since it may have a (single)-black-hole interpretation, as the Pleba\'nski--Demia\'nski solution does. We do not have a definite answer to this yet. On the one hand, the solution does contain various limits which do have Lorentzian sections, such as the Pleba\'nski--Demia\'nski and Kerr-NUT solutions. On the other hand, the solution also contains the triple-collinearly-centered Gibbons--Hawking solution as a special case, which does not admit a Lorentzian section. This fact cannot rule out the existence of a Lorentzian section for our general solution, because a solution may have a Lorentzian section even though one of its subclasses does not: as an example, the Kerr-NUT solution contains the Eguchi--Hanson solution as a special case and admits a Lorentzian section. We remark that if a Lorentzian section could be found, it is natural to interpret the lowest crossed diamond in Fig.~\ref{fig_box1}(b) as the black-hole curvature singularity, hidden behind the horizon at $y=x_1$.

It may also be interesting to study the algebraic properties of the solution. The Pleba\'nski--Demia\'nski solution is algebraically special, and is so far the most general known Type-D solution in four dimensions. Based on the fact that our solution is a generalisation of the Ricci-flat Pleba\'nski--Demia\'nski solution and that it possesses a very compact form, one may suspect that the solution may have some special algebraic properties. Of course, since a Lorentzian section is still lacking, one has to study this in the regime with Euclidean signature.

Recall that the general Pleba\'nski--Demia\'nski solution carries both electric and magnetic charges. A challenging problem is then to seek a generalisation of this solution, by adding electric and/or magnetic charges. It is even more challenging to seek an (A)dS generalisation. We have mentioned that our solution emerges from the same ISM construction as the triple-collinearly-centered Taub-NUT solution. The latter solution has an (A)dS generalisation, which can be obtained as a special case from the general construction carried out by Calderbank and Pedersen \cite{Calderbank:2002}. These facts lead us to believe that it is worth more effort to consider the possibility of these generalisations in the future. One may even try to make guesses, by choosing an ansatz similar in form to our metric.

The new AF instanton was identified within this solution, as mentioned previously. Then a natural question to ask is whether the solution admits more new completely regular gravitational instantons. The analysis is straightforward with the aid of the rod structure presented in this paper, but the actual calculations are rather involved. More careful and exhaustive analysis is needed.

Other possible future directions include the interpretation of the solution when embedded in string theory, in terms of a system of interacting D-branes. More detailed analysis of the solution in Kaluza--Klein theory is also worth pursuing.

\bigbreak\bigskip\bigskip\centerline{{\bf Acknowledgement}}
\nobreak\noindent This work was supported by the Academic Research Fund (WBS No.: R-144-000-333-112) from the National University of Singapore.

\appendix

\newsection{Double-centered and single-centered limits}

\label{section_appendix1}

In this appendix, we consider degenerate limits of the solution (\ref{metric_general}) in which the final solution contains only two or one turning points. The resulting solutions include the Kerr-NUT, the double-centered Taub-NUT, the non-self-dual as well as the self-dual Taub-NUT solutions.

\newsubsection{Kerr-NUT solution}

It is known that a scaling limit \cite{Plebanski:1976gy} of the Pleba\'nski--Demia\'nski solution results in the Kerr-NUT solution (as a subclass of the Carter--Pleba\'nski solution \cite{Carter,Plebanski}). This limit corresponds to sending the first or third turning point to infinity while zooming in to the region around the remaining two turning points. This limit is still present in the general solution (\ref{metric_general}), although in a more subtle way. To recover it, one first defines the parameters
\be
x_1=x_2-2\epsilon\,,\quad x_3=-x_2-c_1\epsilon\,,\quad x_4=-x_2-c_2\epsilon\,,\quad \nu=1+c_3\epsilon/x_2\,,\quad k=l/\epsilon^3,
\ee
and the coordinates $(r,\theta)$ by
\be
x=x_3+\epsilon\left[4 rx_2\sqrt{\frac{-x_2}{l(2+c_1+c_2+2c_3)}}+\frac{(c_1+c_3)(2+c_1+c_3)}{2+c_1+c_2+2c_3}\right],\quad y=x_2-\epsilon(1+\cos\theta)\,.
\ee
After making the substitutions
\ba
\tau&\rightarrow& \sqrt{-\frac{\epsilon(2+c_1+c_2+2c_3)}{x_2}}\tau+\frac{x_2\sqrt{k}(K_2[1]+K_4[1])}{a_4(x_1-x_2)(x_2-x_3)(x_2-x_4)}\phi\,,\quad \cr
\phi&\rightarrow& \frac{2x_2\sqrt{k}}{a_4(x_1-x_2)(x_2-x_3)(x_2-x_4)}\phi\,,
\ea
and taking the limit $\epsilon\rightarrow 0$, one can show that the first turning point is pushed to infinity. The metric in this limit becomes
\ba
\label{Euclidean Kerr-bolt}
{\dif s}^{2}&=&{\frac {{\Delta}}{{\Sigma}}} [{\dif\tau}+ \left( 2n
\cos \theta +a \sin  ^{2} \theta
  \right) {\dif\phi}] ^{2}+{\frac {
  \sin ^{2} \theta  }{{\Sigma}}}[a{\dif\psi}-
 \left( {r}^{2}-{n}^{2}-{a}^{2} \right) {\dif\phi}] ^{2}\cr
 &&+{\Sigma} \left( {\frac {{\dif r}^{2}}{{\Delta}}}
+{{\dif\theta}}^{2} \right),\cr
{\Sigma}&=&{r}^{2}- \left( n-a\cos \theta  \right) ^{
2},\qquad
{\Delta}={r}^{2}-2mr-{a}^{2}+{n}^{2},
\ea
with parameters
\ba
m&=&\frac{\sqrt{l}[(c_1+c_3)^2+(c_2+c_3)^2+2c_1+2c_2+4c_3]}{\sqrt{-64x_2^3(2+c_1+c_2+2c_3)}}\,,\cr
n&=&-\frac{\sqrt{l}[(c_1+c_3)(c_2+c_3)+2+c_1+c_2+2c_3]}{\sqrt{-16x_2^3(2+c_1+c_2+2c_3)}}\,,\cr
a&=&\sqrt{-\frac{l(2+c_1+c_2+2c_3)}{16x_2^3}}\,.
\ea
This is the familiar form of the Kerr-NUT solution.

However, there is a more natural limit in which one can recover the Kerr-NUT solution. As mentioned above, one can join up two adjacent rods to eliminate the turning point between them, instead of sending it to infinity. Without loss of generality, here we consider the limit (\ref{join_rods34}), in which the third rod is joined up with the fourth. We first define $k$ in terms of $l$, and choose $a_4$ appropriately:
\be
\sqrt{k(1-\nu^2)}=\frac{2l(x_1-x_2)(x_1-x_3)}{x_1(x_2-x_3)}\,,\quad a_4=\frac{4l}{(x_1-x_4)(x_2-x_3)}\,.
\ee
We then bring the solution to its natural Killing coordinates by doing the substitutions
\be
\tau\rightarrow \sqrt{1-\nu^2}(\tau-b\phi)\,,\quad \phi\rightarrow -\phi/\sqrt{1-\nu^2}\,,
\ee
with $b$ given by
\be b=\frac{2l[(1-\nu^2)(x_1^2+x_4^2)+2\nu((x_1+x_4)(x_2+x_3)+2x_1x_4)]}{(1-\nu^2)(x_1-x_4)(x_2-x_3)}\,.\ee
By introducing the parameters $m$, $a$ and $n$
\ba
m&=&-\frac{l[(1+\nu^2)(x_2+x_3)+2\nu(x_1+x_4)]}{(1-\nu^2)(x_2-x_3)}\,,\quad a=-\frac{l(x_1-x_4)}{x_2-x_3}\,,\cr
n&=&-\frac{l[(1+\nu^2)(x_1+x_4)+2\nu(x_2+x_3)]}{(1-\nu^2)(x_2-x_3)}\,,
\ea
and the coordinates $r$ and $\theta$
\ba
r&=&\frac{l(x_2+x_3)(x+y)-2(xy+x_2x_3)}{(x_2-x_3)(x-y)}-\frac{l(1+\nu^2)(x_2+x_3)+2\nu(x_1+x_4)}{(1-\nu^2)(x_2-x_3)}\,,\cr
\cos\theta&=&\frac{(x_1+x_4)(x+y)-2(xy+x_1x_4)}{(x_1-x_4)(x-y)}\,,
\ea
the metric is brought to the Kerr-NUT solution in the familiar form (\ref{Euclidean Kerr-bolt}). The corresponding joining-up limit of the Pleba\'nski--Demia\'nski solution results in the double-centered Gibbons--Hawking solution, a subclass of the Kerr-NUT solution with infinite asymptotic NUT charge.

A third way to eliminate a turning point from the rod structure is to set the length of say the second rod to zero, while keeping the third rod finite. In the touching double-Kerr-NUT picture, this corresponds to shrinking the size of the first Kerr-NUT to zero. This limit is achieved by taking
\be
\label{Kerr_NUT_limit3}
a_{3,4}=0\quad {\rm or}\quad a_{0,1}=0\,.
\ee
We will not present the detailed coordinate transformations here.

\newsubsection{Double-centered Taub-NUT solution}

To recover the double-centered Taub-NUT solution, it is convenient to use the natural Killing coordinates $(\tilde{\tau},\tilde{\phi})$ defined in (\ref{natural_killing_vectors}). We define
\be
\nu=1-c_1\epsilon/l\,,\quad k=2c_1l/\epsilon\,,\quad x_3=-x_2(1-\epsilon)\,,\quad x_4=-x_1(1-c_2\epsilon)\,,
\ee
and take the limit $\epsilon\rightarrow 0$. We then introduce the new parameters $(a,n_1,n_2)$ by
\ba
a=\frac{2c_1x_1x_2}{x_1^2-x_2^2}\,,\quad n_1=\frac{l(1+c_2)(x_1+x_2)}{4(x_1-x_2)}\,,\quad n_2=\frac{l(1-c_2)(x_1-x_2)}{4(x_1+x_2)}\,,
\ea
and the coordinates $(r,\theta)$ by
\ba
r\sin\theta&=&\frac{2c_1\sqrt{-(x^2-x_1^2)(x^2-x_2^2)(y^2-x_1^2)(y^2-x_2^2)}}{(x_1^2-x_2^2)(x-y)^2}\,,\cr
r\cos\theta&=&\frac{2c_1(xy-x_1^2)(xy-x_2^2)}{(x_1^2-x_2^2)(x-y)^2}\,.
\ea
After dropping the tildes on the natural Killing coordinates, and taking $\phi\rightarrow -\phi$, the metric is brought to the familiar form:
\ba
{\dif s}^{2}&=& V^{-1} \left( {\dif\tau}+A \right) ^{2}+V
 ( {\dif r}^{2}+{r}^{2} {{\dif\theta}}^{2}+ r^2 \sin ^{2}
 \theta\, {{\dif\phi}}^{2}  )\,,\quad V=1+\sum_{i=1}^2 \frac{2n_i}{r_i}\,,\cr
 A&=& \sum_{i=1}^2 \frac{2 n_i (r \cos \theta-d_i)}{r_i}\,\dif\phi\,,\quad r_i=\sqrt {{r}^{2}+{{d_i}}^{2}-2 {d_i} r \cos \theta}\,,\quad d_{1,2}=\mp a\,.
\ea
The parameter ranges that we are interested in, (\ref{ranges_parameter1}) and (\ref{ranges_parameter2}), imply that $-1< c_2< 1$, which in turn ensures that $n_{1,2}$ are non-negative, consistent with the requirement that the ``magnetic masses'' are non-negative.

\newsubsection{Non-self-dual Taub-NUT solution}

As mentioned in Sec.~5.1, the case
\be
a_{0,4}=0\,,
\ee
describes a (non-self-dual) Taub-NUT touching another Taub-NUT, resulting in a larger Taub-NUT. To map this solution to the familiar form of the Taub-NUT solution, we choose the gauge $a_3=1$ and write $X=(x-x_2)x(x-x_4)$. Note that one has $x_1=-\infty$ and $x_3=0$. We then define
\be
r=\frac{\sqrt{k}(\nu^2 x-y)}{\sqrt{1-\nu^2}(x-y)}\,,\quad \cos\theta=\frac{2(xy+x_2x_4)-(x_2+x_4)(x+y)}{(x_2-x_4)(x-y)}\,,
\ee
and make the substitutions
\be
\tau\rightarrow \sqrt{1-\nu^2}\tau+\frac{2\nu\sqrt{k}(x_2+x_4)}{x_2-x_4}\phi\,,\quad \phi\rightarrow -\frac{2\sqrt{k}}{x_2-x_4}\phi\,.
\ee
The metric then becomes the Taub-NUT solution, in the form (\ref{Euclidean Kerr-bolt}) with
\be
m=\frac{\sqrt{k}(1+\nu^2)}{2\sqrt{1-\nu^2}}\,,\quad n=-\frac{\nu\sqrt{k}}{\sqrt{1-\nu^2}}\,,\quad a=0\,.
\ee

It is interesting to note that, in this limit, $x_{2,4}$ make no appearance in the physical parameters and become redundant. One is thus free to choose any value for $a_{1,2,3}$ without changing the physical space, as long as the solution does not become degenerate. The non-trivial parameters are $k$ and $\nu$, which can be interpreted as the mass and NUT charge parameter respectively. It can be seen that the special case with $a_0=a_4=\nu=0$ is the Schwarzschild solution.

This is also an example of joining up two rods: the second rod is joined up with the third. If instead one insists that $-\infty\le x_4\le x_2$, then one has to close the box by $x=x_4$ (instead of $x=x_1$) as its second rod. In this case, it is the first rod that joins up with the second. The final solution is of course still the non-self-dual Taub-NUT solution.

\newsubsection{Self-dual Taub-NUT solution}

In this subsection, we show how the self-dual Taub-NUT solution is recovered from the NUT-charged C-metric (\ref{metric_NUT_C-metric}). We  first define the parameters and coordinates
\be
x_1=x_2-\epsilon\,,\quad k=l\epsilon^2,\quad x=x_2+w\epsilon\,,\quad y=x_2-z\epsilon\,,
\ee
and make the substitutions
\be
\tau\rightarrow \tau+\frac{\sqrt{k}x_1x_3}{(x_1-x_2)(x_2-x_3)}\phi\,,\quad \phi\rightarrow -\frac{2\sqrt{k}x_2}{(x_1-x_2)(x_2-x_3)}\phi\,.
\ee
After taking the limit $\epsilon\rightarrow 0$, the metric then becomes
\be
{\dif s}^{2}= V^{-1} \left( {\dif\tau}+2n\cos\theta\, \right) ^{2}+V
 ( {\dif r}^{2}+{r}^{2} {{\dif\theta}}^{2}+ r^2 \sin ^{2}
 \theta\, {{\dif\phi}}^{2}  )\,,\quad V=1+\frac{2n}{r}\,,
\ee
if one identifies
\be
r=\frac{\sqrt{l}x_2(z-w-1)}{w+z}\,,\quad \cos\theta=\frac{w^2+w+z^2-z}{w^2+w+z-z^2}\,,\quad n=-\frac{\sqrt{l}x_2x_3}{2(x_2-x_3)}\,.
\ee
This is the standard form of the self-dual Taub-NUT solution.

\bigskip\bigskip

{\renewcommand{\Large}{\normalsize}
}

\end{document}